\documentclass[aps,preprint]{revtex4}%
\usepackage{amsfonts}
\usepackage{amsmath}
\usepackage{amssymb}
\usepackage{graphicx}%
\providecommand{\U}[1]{\protect\rule{.1in}{.1in}}
\begin{document}
\author{}
\title{Entropy spectrum of the apparent horizon of Vaidya black holes via adiabatic invariance}
\author{Ge-Rui Chen}
\email{chengerui@emails.bjut.edu.cn}
\author{Yong-Chang Huang}
%\email{ ychuang@bjut.edu.cn}
\affiliation{Institute of Theoretical Physics, Beijing University of
Technology, Beijing, 100124, China}

\begin{abstract}

The spectroscopy of the apparent horizon of Vaidya black holes is
investigated via adiabatic invariance. We obtain an equally spaced
entropy spectrum with its quantum equal to the one given by
Bekenstein \cite{jdb}. We demonstrate that the quantization of entropy and area
is a generic property of horizon, not only for stationary black holes, and the results also exit in a dynamical black
hole.  Our work also shows that the quantization of black hole is closely related to Hawking temperature, which is an interesting thing.

\end{abstract}

\maketitle

Since the first exact solution of Einstein equation was found out,
studying black holes' properties has become an important part of
gravitational physics. With the discovery of laws of black hole mechanics \cite{hk1,jmb,jdb} and Hawking radiation \cite{hk2,hk3},
laws of black hole thermodynamics have been built up successfully,
which causes deep, unsuspected connections among classical general
relativity, quantum physics and statistical mechanics. Since
thermodynamics is a phenomenological theory, there should exists a
more fundamental theory of gravity just as statistical mechanics,
then the statistic origin of black hole entropy becomes an
interesting problem. In 1970' s, Bekenstein proved that the quantum
of the black hole horizon is given as $(\Delta A)_{min}=8\pi l_p^2$
\cite{jdb}, and he also showed that the horizon area can be treated
as an adiabatic invariance.  In 1998, Hod \cite{sh} proposed the area
spectrum $(\Delta A)_{min}=4\ln3 l_P^2$ by employing the real part
of quasinormal frequencies of a black hole and Bohr' s correspondence
principle. Later on, based on the proposal of adiabaticity of the
horizon area and quasinormal frequencies, Kunstatter \cite{gk} got
the entropy spectrum of a D-dimensional black hole which is the same
as Hod' s result. In 2008, Maggiore \cite{mm} found that when a
classical black hole is perturbed, its relaxation is governed by a
set of quasinormal modes with complex frequencies
$\omega=\omega_R+i\omega_i$ whose behavior is the same as that of
damped harmonic oscillators with real frequencies
$(\omega_R^2+\omega_I^2)^\frac{1}{2}$, rather than simply
$\omega_R$. Adapting the same derivation by Hod, he found that the
area of the horizon of a Schwarzschild black hole is quantized in
units $(\Delta A)_{min}=8\pi l_p^2$, which is the same as
Bekenstein' s original result. Recently, Majhi and Vagenas \cite{bec}
proposed a new approach to derive the entropy spectrum and the
horizon area quantum utilizing solely the adiabaticity of black
holes and the Bohr-Sommerfeld quantization rule. Later on there were
many works using this method to investigate the entropy spectrum of
different kinds of stationary black holes \cite{dyc1,xxz,lxm,hll3,qqj,czl,hll,dyc,rtvck,jv,ssvck,dyc2,hll2,hll9,swz}.
We think that this method is closely related to Parikh and Wilczek' s tunneling
method \cite{mk1,mk2}, so it should be applied to broader
circumstances. In this paper, we extend Majhi and Vagenas' method to investigate the spectroscopy of the apparent horizon of Vaidya black holes. We demonstrate that the quantization of entropy and area is a generic property of horizon, not only for stationary black holes, and the results also exit in a dynamical black
hole.  Our work also shows that the quantization of black hole is closely related to Hawking temperature, which is an interesting thing.

The line element of the Vaidya black hole is given by
\begin{eqnarray}
ds^2=-[1-\frac{2M(\upsilon)}{r}]d\upsilon^2+2d\upsilon
dr+r^2(d\theta^2+\sin^2\theta d\varphi^2)\label{rn5},
\end{eqnarray}
where $M(\upsilon)$ is the mass of the black hole and $\upsilon$ is
the advanced Eddington time coordinate. From
$g_{\upsilon\upsilon}=0$, we get the apparent horizon
$r_{AH}=2M(\upsilon)$ which is also time-like limit surface. Refs.
\cite{zfn,xl} calculated the Hawking temperature of the apparent
horizon using tunneling method and Damour-Ruffini method, and
established the thermodynamics for the apparent horizon of Vaidya
black holes as follows
\begin{eqnarray}
dM(\upsilon)=T_{AH}dS_{AH},\label{rn8}
\end{eqnarray}
where $T_{AH}=\frac{1}{8\pi M(\nu)}$ and $S_{AH}=4\pi M^2$.

Let us consider the quantization of the apparent horizon of Vaidya
black holes by applying Majhi and Vagenas' adiabatic invariance
method \cite{bec}. Consider an adiabatic invariant quantity
\begin{equation}
I=\int p_i
dq_i=\int\int^{p_i}_{0}dp'_{i}dq_{i}=\int\int_0^H\frac{dH'}{\dot{q_i}}dq_i=
\int\int_0^HdH'd\tau+\int\int_0^H\frac{dH'}{\dot{r}}dr\label{rn3},
\end{equation}
where $p_i$ is the conjugate momentum of the coordinate $q_i$ with
$i=0,1$ for which  $q_0=\tau$ and $q_1=r$. Note that we use
the Euclidean time $q_0=\tau$ and the Einstein summation convention.
To get the third equation, we have used Hamiltion canonical equation
$\dot{q_i}=\frac{dH}{dp_i}$, where the Hamiltonian $H$ is the total
energy of the black hole. In order to calculate the adiabatic
invariant quantity we shall obtain the quantity $\dot{r}$ that
appears in Eq. (\ref{rn3}). As Ref. \cite{bec}, let us consider the
radial null paths. Our subsequent analysis will concentrate on the
outgoing paths, since these are the ones related to the quantum
mechanically nontrivial features \cite{mk1}. Because $\tau$ is the
Euclidean time, like Ref. \cite{bec}, we use the transformation
$\upsilon\rightarrow i\tau$ to Euclideanize the metric (\ref{rn5})
and get the radial null paths,
\begin{eqnarray}
\dot{r}\equiv
\frac{dr}{d\tau}=\frac{i}{2}[1-\frac{2M}{r}]=R_+(r).\label{rn4}
\end{eqnarray}
Now, using Eq. (\ref{rn4}), we have
\begin{eqnarray}
\int\int_0^HdH'd\tau=\int\int_0^HdH'\frac{dr}{R_+(r)}=\int\int_0^HdH'\frac{dr}{\dot{r}},
\end{eqnarray}
and the adiabatic invariant quantity (\ref{rn3}) reads
\begin{eqnarray}
I=\int p_i
dq_i=2\int\int_0^HdH'd\tau=2\int\int_0^HdH'\frac{dr}{\dot{r}}.
\end{eqnarray}
In Ref. \cite{bec}, the authors perform the $\tau$-integration by
considering the periodicity of imaginary time $\tau$ for static
black holes, however we do not know if the periodicity is valid for
dynamical black holes. Fortunately, we can do the $r$-integration. This is the main difference between Ref. \cite{bec} and this work.
Using the technology in Parikh and Wilczek' s tunneling method
\cite{mk1,mk2}, we can get
\begin{eqnarray}
\int\int_0^HdH'\frac{dr}{\dot{r}}&=&\int\int_{r_{in}}^{r_{out}}\frac{dr}{\frac{i}{2}(1-\frac{2M}{r})}dH'=\int\int_{r_{in}}^{r_{out}}\frac{2rdr}{i(r-2M)}dH' \\
&=&\int_0^H4\pi M dH'=\pi\int_0^H\frac{dH'}{\kappa},
\end{eqnarray}
where $\kappa=\frac{1}{4M}$ is the surface gravity of the apparent
horizon. So we obtain the adiabatic invariance,
\begin{eqnarray}
I=\int p_i
dq_i=2\pi\int_0^H\frac{dH'}{\kappa}=\hbar\int_{0}^{H}\frac{dH'}{T_{AH}}=\hbar
S_{AH},
\end{eqnarray}
where we have used the temperature of the apparent horizon of Vaidya
black holes $T_{AH}=\frac{\hbar \kappa}{2\pi}$ and the first law of
thermodynamics established on the apparent horizon (\ref{rn8}). At
last, implementing the Bohr-Sommerfeld quantization rule,
\begin{eqnarray}
\int p_i dq_i=nh,
\end{eqnarray}
we derive the entropy spectrum,
\begin{eqnarray}
S_{AH}=2\pi n,
\end{eqnarray}
where $n=1,2,3,...$, and it is straightforward to see that the
spacing in the entropy is given by
\begin{eqnarray}
\Delta S_{AH}=S_{(n+1)AH}-S_{(n)AH}=2\pi\label{rn7}.
\end{eqnarray}
Thus, the entropy spectrum is quantized and equidistant for the
apparent horizon. Recalling that in the framework of Einstein's
theory of gravity, black hole entropy is proportional to the black
hole horizon area \cite{jdb}, $S_{AH}=\frac{A}{4 l_{p}^2}$. It is
evident that if we employ the spacing of the entropy spectrum given
in Eqs. (\ref{rn7}), the quantum of the apparent horizon area has
the form,
\begin{eqnarray}
\Delta A=8\pi l_p^2\ ,
\end{eqnarray}
which is the same as the area quantum derived by Bekenstein
\cite{jdb}.

Let us give some discussions. We considers a tunneling process and uses the
technology of the tunneling method \cite{mk1,mk2} which has been successfully used to calculate Hawking temperature of a variety of spacetimes, so the results of the quantization of black hole have more generality. Furthermore, our work also shows that there exists close relationship between the entropy quantum and Hawking temperature, which reflects to some extent the viewpoint of the emergent perspective of gravity that temperature means existence of underlying degrees of freedom \cite{tp}. At last, from the calculation, it is easy to see that the ingoing path does not contribute to the adiabatic invariant quantity (\ref{rn3}), so we only need to consider the outgoing paths.

In summary, we investigate the spectroscopy of the apparent horizon of
a Vaidya black hole via adiabatic invariance and Bohr-Sommerfeld
quantization rule, and obtain the quantized entropy and area
spectrum which are in accordance with Bekenstein' s original results \cite{jdb}. These
results indicate that the quantization of entropy and area of the
black hole horizon is a generic property of horizon, not only for
stationary black holes, and the quantization of black hole is closely related to Hawking temperature.

\begin{acknowledgements}
This work is supported by National Natural Science Foundation of
China (No. 11275017 and No. 11173028).
\end{acknowledgements}

\end{document}